\documentclass[twocolumn,preprintnumbers,amsmath,amssymb]{revtex4}

\usepackage{graphicx}
\usepackage{dcolumn}
\usepackage{bm}

\begin{document}


\title{Quantum Measurements and the $\kappa$--Poincar\'e Group}
\author{Abel Camacho}
\email{acq@xanum.uam.mx}
\affiliation{Department of Physics \\
Universidad Aut\'onoma Metropolitana--Iztapalapa\\
Apartado Postal 55--534, C.P. 09340, M\'exico, D.F., M\'exico.}

\author{ A. Camacho--Galv\'an}
\email{abel@servidor.unam.mx}
\affiliation{ DEP--Facultad de Ingenier{\'\i}a\\
Universidad Nacional Aut\'onoma de M\'exico.}

\date{\today}
             \begin{abstract}
The possible description of the vacuum of quantum gravity through
the so called $\kappa$--Poincar\'e group is analyzed considering
some of the consequences of this symmetry in the path integral
formulation of nonrelativistic quantum theory. This study is
carried out with two cases, firstly, a free particle, and finally,
the situation of a particle immersed in a homogeneous
gravitational field. It will be shown that the
$\kappa$--Poincar\'e group implies the loss of some of the basic
properties associated to Feynman's path integral. For instance,
loss of the group characteristic related to the time dependence of
the evolution operator, or the breakdown of the composition law
for amplitudes of events occurring successively in time.
Additionally some similarities between the present idea and the so
called restricted path integral formalism will be underlined.
These analogies advocate the claim that if the
$\kappa$--Poincar\'e group contains some of the physical
information of the quantum gravity vacuum, then this va\-cuum
could entail decoherence. This last result will also allow us to
consider the possi\-bility of analyzing the continuous measurement
problem of quantum theory from a group--theoretical point of view,
but now taking into account the $\kappa$--Poincar\'e symmetries.
\bigskip

\end{abstract}
\maketitle

\section{Introduction}
\bigskip

\cite{[6]}

The analysis of this large distance physics could begin with the
$\kappa$--Poincar\'e group \cite{[4]}, since it could contain some
of the physical features in connection with the quantum gravity
vacuum \cite{[5]}. The research in this direction has been,
mainly, focused on the modifications of the Heisenberg algebra and
on the measurability of distances, that this symmetry could
render. In other words, the consequences of this group upon
non--relativistic physics has been, as far as the authors know,
not thoroughly, analyzed. Clearly the expected deviations from the
usual behavior must be (as in other cases, for instance,
gravity--wave interferometers \cite{[2]}) very small.

Nevertheless, the analysis of the effects, upon non--relativistic
theory, of the $\kappa$--Poincar\'e group could render important
conclusions. Indeed, not only, some of the physics in the scale of
low velocities could be understood, but also a second topic of
current interest  could obtain benefits from the present analysis,
namely, the so called continuous quantum measurement problem
\cite{[6]}. In other words, we look for the possible description
of the non--relativistic continuous quantum measurement pro\-blem
employing the $\kappa$--Poincar\'e group. The analysis of the
group--theoretical approach to this problem has already been
considered \cite{[6]}, but clearly more work in this direction is
needed.

We may justify this last possibility noting that the
quantum--gravity--induced loss of quantum coherence is one of the
main alternative viewpoints on quantum gra\-vity \cite{[7]}.
Additionally, remembering that decoherence is related to quantum
measurement \cite{[8]}, we may pose the fo\-llowing question: if
the $\kappa$--Poincar\'e group describes some of the physical
features involved in the quantum deformations of relativistic
symmetries (induced by the quantum gravity vacuum \cite{[2]} and
if these quantum deformations imply that the measurability of
distances is bounded by a root-mean square deviation that grows
with time \cite{[9]} (a fact that comprises the loss of quantum
coherence \cite{[2]}, and hence related to quantum measurement),
could this symmetry be used as part of the group--theoretical
framework co\-nnected with the description of the continuous
quantum measurement problem?

In the present work we analyze the consequences of the
$\kappa$--Poincar\'e group upon the path integral of a
non--relativistic particle. In particular two cases will be
consi\-dered, a free particle, and finally, a particle immersed in
a homogeneous gravitational field. It will be shown that though,
for practical purposes, the deviations from the usual dynamics are
quite small, the effects of this sy\-mmetry might have
far--reaching consequences. Indeed, the group property of the
evolution operator, concerning the time parameter \cite{[10]}, is
lost. Furthermore, the composition law for amplitudes of events
occurring successively in time \cite{[10]} breaks down.

Additionally, it will be shown that some of the features related
with the so called Restricted Path Integral Formalism (RPIF)
\cite{[6]} do appear in connection with the present approach.
Particularly, it will become clear that the present model contains
(similarly to the predictions of RPIF for the case of a particle
whose position is being monitored) a quantum corridor, the one
contains all the possible trajectories allowed by our starting
symmetries. This last result advocates the possibility of
analyzing the continuous quantum measurement problem using the
$\kappa$--Poincar\'e group as part of the involved symmetries.
\bigskip

\section{Path Integrals and $\kappa$--Poincar\'e Group}
\bigskip

\subsection{Minimal length and time in non--relativistic path integrals}
\bigskip

As mentioned in the previous section, our starting point will be
the modified Heisenberg algebra encoded in the
$\kappa$--Poincar\'e group \cite{[4], [5]}. In particular we will
consider one of the possibilities that this symmetry offers,
namely

\begin{eqnarray}
\Delta\hat t\Delta\hat x_k \geq {\hbar \over 2\kappa
c^2}\vert<\hat x_k>\vert, \label{GroupI}
\end{eqnarray}

\begin{eqnarray}
\Delta\hat p_l\Delta\hat x_k \geq {\hbar \over 2}\delta_{lk},
\label{GroupII}
\end{eqnarray}

\begin{eqnarray}
 \Delta\hat E\Delta\hat p_k \geq {\hbar \over 2}, \label{GroupIII}
\end{eqnarray}

\begin{eqnarray}
\Delta\hat t\Delta\hat p_k = {\hbar \over 2\kappa c^2}\vert<\hat
p_k>\vert. \label{GroupIV}
\end{eqnarray}

The question concerning path integrals involves the modifications,
if any, that (\ref{GroupI}) could render. Looking for the
transition amplitude of finding, at time $t_b$, the particle at
point $x_b$ (here for the sake of clarity we restrict the
situation to one--dimensional motion), knowing that at time $t_a$
it was located at point $x_a$ we find \cite{[9], [10]}

\begin{eqnarray}
A = \int_{x_a}^{x_b}
Dx\exp\Bigl\{{i\over\hbar}\int_{t_a}^{t_b}\Bigl[{M\over 2}\dot x^2
+ V(x)\Bigr]\Bigr\}. \label{AmpI}
\end{eqnarray}

The last path integral is calculated considering the limit
\cite{[9], [10]} when $N\rightarrow \infty$ of

\begin{eqnarray}
A \cong {1\over\sqrt{2i\pi\hbar\epsilon/M}}
\Bigl[\prod_{n=1}^{n=N}\int_{-\infty}^{\infty}{dx_n\over\sqrt{2i\pi\hbar\epsilon/M}}
\Bigr]\nonumber \\
\times\exp\Bigl\{{i\epsilon\over\hbar}\sum_{n=1}^{N+1}\Bigl[{M\over
2}({x_n - x_{n-1}\over\epsilon})^2 + V(x_n)\Bigr]\Bigr\}.
\label{AmpII}
\end{eqnarray}

Here $\epsilon = (t_b - t_a)/N$, and
$N\rightarrow\infty\Rightarrow\epsilon\rightarrow 0$.

Clearly, the step that takes us from (\ref{AmpII}) to (\ref{AmpI})
entails the fact that we may, without any restriction, consider,
simultaneously, the distance differences $x_n - x_{n-1}$, and the
time difference $\epsilon$, as small as we wish.

This seems to contradict (\ref{GroupI}). A possible solution to
this problem considers the introduction in the path integral of a
minimal length, and also of a minimal time, i.e., the limit
$N\rightarrow\infty\Rightarrow\epsilon\rightarrow 0$ is now
forbidden, since it implies the simultaneous vanishing of time and
length differences. Of course, a second possibility is that
Feynman's formulation becomes invalid for distances $\vert x_a -
x_b\vert$ smaller than a certain value, and in consequence
(\ref{AmpII}) becomes meaningless.

Nevertheless, here the idea is to consider the first option and
try to find the kind of modifications that, in the
non--relativistic level of quantum theory, the
$\kappa$--Poincar\'e group might render. We may justify the
introduction of (\ref{GroupI}), in the calculation of the path
integral, noting that in Feynman's formulation any trajectory (not
only the classical one), joining the corresponding endpoints, is a
possible trajectory for the motion of the involved particle
\cite{[9], [10]}. From this last argument it is easier to
understand the need of considering the effects, upon the motion
possibilities, that this restriction imposes. Otherwise, if we
believe in the fact that the particle may move along any
trajectory appearing in the path integral, then the particle could
end up moving without taking into account the main symmetries of
the situation.
\bigskip

Therefore, here we will assume:
\bigskip

(i) For a certain distance $\vert x_b - x_a\vert$, and a certain time difference $t_b - t_a$, there exists a natural
number $N$, such that we may define $\epsilon \cong (t_b - t_a)/(N + 1)$, and $l_m \cong \vert x_b - x_a\vert/N =
\vert x_n - x_{n-1}\vert, \forall n\in\Bigl\{1, 2, 3, ...N+ 1\Bigr\}$, with $\epsilon l_m\geq {\hbar \over 2\kappa c^2}
\vert<\hat x_k>\vert$.
\bigskip

(ii) In the functional integration related to (\ref{AmpII}) we are
now not allowed to take the limit $N\rightarrow\infty$. In other
words, the largest value of $n$ in (\ref{AmpII}) is determined by
(i), i.e., $N$.
\bigskip

Though here we have a formalism very similar to Feynman's one, in
our approach the introduction of (\ref{GroupI}) discards the limit
$N\rightarrow\infty$. The consequences of these two conditions
will become clear in the next two subsections.
\bigskip

\subsection{Free particle}
\bigskip

Let us now consider the simplest case,  the path integral for a
one--dimensional free particle. Here (i) and (ii) mean that, for
instance, if we consider the first Riemann integral, associated to
$x_1$, we now are not allowed to integrate from ${-\infty}$ to
${\infty}$, since this would imply considering values of $x_1$
such that $\vert x_1 - (x_a + x_2)/2\vert <l_m$, a fact that
violates (i). Hence the integration limits, associated to $x_1$,
run from ${-\infty}$ to $(x_a + x_2)/2 - l_m$ and from $(x_a +
x_2)/2 + l_m$ to ${\infty}$. In other words, those values of
$x_1$, lying within the interval $[(x_a + x_2)/2 - l_m, (x_a +
x_2)/2 + l_m]$, are not considered in the functional integration.
This argument is used in the $N$ integrals connected with
(\ref{AmpII}). Performing the integrations we find

\begin{eqnarray} A_{(f)} = \sqrt{{M\over2i\pi\hbar(t_b - t_a)}}
\exp\Bigl\{{iM\over 2\hbar(t_b- t_a)}(x_b- x_a)^2\Bigr\}\nonumber \\
\times\Bigl[1 - \sqrt{{2l_mv_m\over i\pi\lambda c}}F({il_mv_m\over
\lambda c})\Bigr]\Bigl[1 - \sqrt{{3l_mv_m\over 2i\pi\lambda
c}}F({3il_mv_m\over 4\lambda c})\Bigr]\nonumber\\
\times...\times\Bigl[1 - \sqrt{{(N+1)l_mv_m\over Ni\pi\lambda
c}}F({(N+1)il_mv_m\over 2N\lambda c})\Bigr]. \label{AmpIII}
\end{eqnarray}
\bigskip

In (\ref{AmpIII}) $\lambda$ denotes the Compton wavelength of the
involved particle, $v_m = l_m/\epsilon$, while $F(z)$ represents
the hypergeometric function $H(\alpha;\beta;z)$ \cite{[11]}, with
the conditions $\alpha = 1/2$ and $\beta = 3/2$.
\bigskip

\subsection{Homogeneous gravitational field}
\bigskip

Let us now consider a more realistic case, a particle immersed in a homogeneous gravitational field described by the
constant $g$. Proceeding as before we find

\begin{eqnarray} A_{(g)} = A_{(f)}\exp\Bigl\{{i\over\hbar}\Bigl[-{(Mg)^2\over 24M}(t_b- t_a)^3 {4N+3\over 4N+4}
\nonumber\\
-{Mg\over 2}(t_b- t_a)(x_b + x_a){N+2\over N+1}\Bigr]\Bigr\}.
\label{AmpIV}
\end{eqnarray}
\bigskip

\subsection{Minimal length and Feynman's formulation}
\bigskip

Let us now analyze the case in which the minimal length $l_m$
vanishes. In order to do this remember \cite{[11]} that if we
introduce the limit $l_m\rightarrow 0$, then $F({(k+1)il_mv_m\over
2k\lambda c})\rightarrow 1$. This last fact implies that in
(\ref{AmpIII}) we do recover the usual expression for a free
particle \cite{[9], [10]}. In the second case we also recover the
path integral that stems from Feynman's idea, see equation (16.64)
of \cite{[12]}. In other words, the present idea contains the
usual results as a limit case, when there is no minimal length.

If at this point we assume $\epsilon \sim t_p$, and $l_m \sim l_p$
(here $l_p$ and $t_p$ denote the Planck length and time), then
$v_m \sim c$. For an electron we obtain ${l_mv_m\over\lambda c}
\sim 10^{-22}$, and hence $[1 - \sqrt{{2l_mv_m\over i\pi\lambda
c}}F({il_mv_m\over \lambda c})\Bigr]\cong 1 - 10^{-22}/\sqrt{2} +
i10^{-22}/\sqrt{2}$. In the case when $\vert x_a - x_b\vert\sim$ 1
cm, we have that $N\sim 10^{33}$, and hence the possibility of
distinguishing (\ref{AmpIII}) or (\ref{AmpIV}) from Feynman's case
lies outside the current technological capabilities.
\bigskip

\section{Conclusions}

In the present work we have analyzed some of the consequences of
the $\kappa$--Poincar\'e group upon the path integral of a
non--relativistic particle. In particular two cases were
considered, a free particle, and finally, a particle immersed in a
homogeneous gravitational field. The e\-ffects of the
aforementioned symmetry were incorporated introducing, in the path
integral, a minimal length, $l_m$, and a minimal time, $\epsilon$.
The consequences of these two conditions were that now we are not
allowed to integrate over all the configuration space.
Additionally, the usual limit, $N\rightarrow\infty$, present in
Feynman's formulation, is now forbidden.

One of the fundamental properties of Feynman's formulation
comprises the composition law for amplitudes of events occurring
successively in time \cite{[9], [10]}. If $A(x_b; t_b\vert x_a;
t_a)$ denotes the transition amplitude, then this composition law
states \cite{[10]}

\begin{eqnarray}
A(x_b; t_b\vert x_a; t_a) = \int A(x_b; t_b\vert x_c; t_c)A(x_c;
t_c\vert x_a; t_a)dx_c.\label{AmpV}
\end{eqnarray}
\bigskip

Here we have that $t_c\in [t_a, t_b]$. Nevertheless, within the
context of the $\kappa$--Poincar\'e group this can not be valid,
since the presence of a non--vanishing $l_m$ implies that now it
is not allowed to integrate over the whole configuration space.

The group property, related to the time dependence, of the
evolution operator, $\hat U(t_b;t_a)$, is a very known fact
\cite{[13]}, a feature that mathematically reads

\begin{eqnarray} \hat U(t_b;t_a) = \hat U(t_b;t_c)\hat
U(t_c;t_a),\label{EvoI}
\end{eqnarray}
\bigskip

\noindent where $t_c\in [t_b, t_a]$.

A fleeting glimpse at the usual definition of evolution operator
\cite{[13]} shows us that it would also be affected by $\Delta\hat
t\Delta\hat x_k\not =0$

\begin{eqnarray} \hat U(t_b;t_a) =
\exp\Bigl\{-{i\over\hbar}\int_{t_a}^{t_b}\Bigl[{\hat p^2\over2m} +
V(\hat x)\Bigr]dt\Bigr\}.\label{EvoII}
\end{eqnarray}
\bigskip

Indeed, if we consider in (\ref{EvoII}) the case in which $t_b =
t_a + \epsilon/2$, then we see that this group property breaks
down, otherwise in the evolution operator $\hat U(t_b;t_a)$ we
would violate (i) and (ii), since we could then ask for the
transition probability associated with $\vert x_b - x_a\vert=
l_m$, and hence $(t_b - t_a)\vert x_b - x_a\vert < {\hbar \over
2\kappa c^2}\vert<\hat x_k>\vert$. In other words, the group
property associated to the time dependence of the evolution
operator disappears with the case $\Delta\hat t\Delta\hat x_k\not
=0$. The extension from the one-dimensional case to a
three-dimensional one is straightforward.

One of the solutions put forward in connection with the continuous
quantum  measurement problem comprises RPIF \cite{[6],[8]}. In
this approach, for the case of a particle whose position is being
monitored, the main idea is the breakdown of the equiprobability
of all the possible trajectories, as a consequence of the presence
of the measuring device. This last fact renders the emergence of a
quantum corridor, which contains all the possible trajectories
that match with the measurement readout. Going back to
(\ref{AmpIII}) (and keeping only the lowest order in
${l_mv_m\over\lambda c}$) it is readily seen that

\begin{eqnarray} \vert A(x_b; t_b\vert x_a; t_a)\vert^2 =
{M\over2\pi\hbar T}\Bigl[1 - 2\sqrt{{l_mv_m\over\pi\lambda c}}
\nonumber\\- 2\sqrt{{3l_mv_m\over 4\pi\lambda c}}-...-
2\sqrt{{(N+1)l_mv_m\over 2N\pi\lambda c}}\Bigr].\label{AmpV}
\end{eqnarray}
\bigskip

 We know that in the case $l_m =0$ (\ref{AmpV}) entails that
all the trajectories have the same probability, though, clearly, a
different phase \cite{[9],[10]}. Consider now any curve $x(t)$
joining $x_a$ and $x_b$. Draw at the initial point, $x_a$, the
straight line associated with velocity $v_m$. From our previous
arguments (see (i) above) it is readily seen that any trajectory
with initial point $x_a$ and lying above $v_m$ is not allowed.
Indeed, any trajectory, joining $x_a$ and $x_b$, that contains
speed values greater than $v_m$ defines a motion such that in a
certain portion of it a distance $l_m$ is travelled in a time
$\Delta t < \epsilon$, and in consequence it violates (i). We may
rephrase this stating that the $\kappa$--Poincar\'e group discards
some trajectories, which in the usual theory have to be included.

A fleeting glimpse at the current literature \cite{[14]}, in
co\-nnection with the quantum measurement problem, readi\-ly shows
that this aforementioned trait also appears in RPIF. This last
fact allows us to consider the possi\-bility of analyzing the
continuous quantum measurement pro\-blem from a group--theoretical
point of view, but now taking into account the
$\kappa$--Poincar\'e group as part of the involved symmetries.
Group methods have already been used in the study of the
measurement problem \cite{[6]}, nevertheless more work in this
context is needed. The results in this direction will be published
elsewhere

Finally, since our approach breaks down the equipro\-bability of
all the trajectories, we may understand the extra terms in
(\ref{AmpV}) (those depending upon $l_mv_m/(\lambda c)$) as
manifestation of decoherence, induced by the breakdown of the
Poincar\'e symmetry. Therefore, we claim that if the vacuum of
quantum gravity is described by this new symmetry \cite{[5]}, then
it might render decoherence.
\bigskip

\Large{\bf Acknowledgments.}\normalsize
\bigskip

We dedicate the present work to Alberto Garc\'{\i}a on the
occasion of his $60^{th}$ birthday. This research was partially
supported by CONACYT Grant 42191--F. A.C. would like to thank A.A.
Cuevas--Sosa for useful discussions and literature hints.

\end{document}